\documentclass[12pt]{article} 
\title{Spacetime is for SU(2)}  
\author{{\it Richard Shurtleff~}\thanks{affiliation and mailing 
address: Department of Applied Mathematics and Sciences, 
Wentworth Institute of Technology, 550 Huntington Avenue, 
Boston, MA, USA, ZIP 02115, telephone number: (617) 989-4338, fax 
number: (617) 989-4591 , e-mail address: shurtleffr@wit.edu}} 
\begin{document} 
          
\maketitle 

\begin{abstract} 

The generators of rotations, boosts and translations are built up based on the commutation and anticommutation relations of the fundamental two dimensional representation of SU(2). Rotations in spacetime derive from the commutation relations of SU(2). Boosts derive from the anticommutation relations of SU(2).

\vspace{0.5cm}
 
\vspace{0.5cm}


\end{abstract}

\pagebreak

\section{Introduction} \label{intro}

We take the group SU(2) to be fundamental and derive spacetime cookbook style, step-by-step. One motivation for finding a recipe for spacetime given SU(2) is to obtain a `spacetime' for SU(3) and other SU($N$)s.

Section 2 sets the notation and recalls some well-known properties of SU(2). SU(2) and the rotation group in 3-space are closely related; they share the same Lie algebra. The distinctions are not emphasized and we concentrate on the Lie algebra. The generators $J$ of SU(2) are called `angular momentum matrices' and SU(2) transformations are called `rotations'.

Section 3 steps through the procedure. Besides angular momentum matrices $J$, we define and discuss other matrices such as `boosts' $K$, `vector matrices' $V$ and `momentum matrices' $P$. Momentum matrices are vector matrices that commute. The  angular momentum matrices $J$ are considered as known in the many representations of SU(2), whereas the $K$s, $V$s, and $P$s are free to define as we wish.  

In Step 1 in the 2-rep of SU(2), we define boost matrices $K_{(2)}$ and vector matrices $V_{(2)}$ as scalar multiples of the $J_{(2)}$s. The boosts $K_{(2)}$ and the $J_{(2)}$s obey the Lie algebra of the Lorentz group of rotations and boosts. No translations yet.

In Step 2, the 4-dimensional  $2\oplus 2$-rep of SU(2) allows boosts $K_{(2\oplus 2)}$ and vector matrices $V_{(2\oplus 2)}$ that with the $J_{(2\oplus 2)}$s satisfy the Lie algebra of the Poincar\'{e} group of rotations, boosts and translations. The $4\times4$ vector matrices $V_{(2\oplus 2)}$ are the `Dirac gamma matrices'.

Momentum matrices $P_{(2\oplus 2)}$ are vector matrices $V_{(2\oplus 2)}$ that are forced to commute. One finds that there are two sets of momentum matrices denoted $P_{\pm\,(2\oplus2)}.$ Each set obeys the Poincar\'{e} algebra.

At the end of Step 2, we have generators that satisfy the Poincar\'{e} algebra, but we do not have the representation for spacetime.

In Step 3 we get spacetime boosts and rotations. Rather than the $2\oplus 2$-rep matrices, we make boosts $K^{i}_{(4)}$ and angular momentum matrices $J^{i}_{(4)}$ from coefficients in the commutation relations with vector matrices. The generators of rotations and boosts in spacetime are the coefficients $a$ and $b$ of $V_{(2\oplus 2)}$  in the commutation relations $[V_{(2\oplus 2)},J_{(2\oplus 2)}]$ = $aV_{(2\oplus 2)}$ and $[V_{(2\oplus 2)},K_{(2\oplus 2)}]$ = $bV_{(2\oplus 2)}.$ The coefficients $a$ and $b$ each has three indices, e.g. $a^{\mu i \nu}$ and $b^{\mu i \nu}$, and form angular momentum matrices ${J^{i}_{(4)}}_{\mu \nu}$ = $a^{\mu i \nu}$ and boost matrices ${K^{i}_{(4)}}_{\mu \nu}$ = $b^{\mu i \nu}.$ These matrices represent rotations and boosts in spacetime.

The label `$(4)$' indicates the $(3\oplus1)$-rep of rotations, i.e. the `4-vector' rep, generated by the new angular momentum matrices. Throughout this paper, no matter what Lie algebra is discussed, the subscript indicates the representation of the SU(2) subalgebra.

At the end of Step 3 we have boosts and rotations for spacetime satisfying the Lorentz algebra. By applying rotations and boosts only to vectors and tensors made from multiples of coordinate differences, invariance under translations is guaranteed and the full Poincar\'{e} algebra is obtained. The generators produce the usual boosts, rotations and translations of spacetime.

The process shows that the 4-vector rep matrices $J^{i}_{(4)}$s descend from coefficients $a$ which originate in the $[J_{(2)},J_{(2)}]$ {\it{commutation}} relations of the fundamental 2-rep of SU(2). And the $K^{i}_{(4)}$s come from the coefficients $b$ from the $\{J_{(2)},J_{(2)}\}$ {\it{anticommutation}} relations of the fundamental 2-rep of SU(2).

Step 4 considers invariants. We show that the invariance of the square of the distance in 3-space ${x^{i}}^2$ follows from the $ij$ antisymmetry of the fundamental commutation relations $[J^{i}_{(2)},J^{j}_{(2)}]$ of SU(2). The invariance of distance under rotations is termed {\it{robust}} because it depends on symmetry.

In contrast, the invariance of the square of the spacetime interval ${x^{i}}^2 - {x^{4}}^2$ for boosts is {\it{not robust.}} This invariance depends on the specific values of the boost matrices $K^{i}_{(4)},$ which are especially simple because the anticommutation relations $\{J_{(2)},J_{(2)}\}$ are simple. The SU(2) anticommutator does not have a sum over $J_{(2)}$s as in SU($N$), $N>2.$ From this point of view the invariance of the spacetime interval under boosts is a coincidence based on the simplicity of SU(2).

\section{The Fundamental SU(2) Rep} \label{SU2}

Unitary $2\times2$ matrices with determinant equal to one form a group under matrix multiplication. A unitary  $2\times2$ matrix $D(\theta)$ can be written as a function of three real parameters $\theta_{i},$ $i \in \{1,2,3\},$ with three generators $J^{i}_{(2)},$
$D_{(2)}(\theta)  \equiv$ $\exp{(i\theta_{i}J^{i}_{(2)})}.$ The `$(2)$' indicates the two-dimensional rep of SU(2). The $J_{(2)}$s must be hermitian and traceless. The $J_{(2)}$s are called `angular momentum matrices' and the $\theta$ parameters are called `angles'. The matrix $D_{(2)}(\theta)$ is a `rotation matrix' that, when applied to a 2-vector, yields a rotated 2-vector.\cite{Tung,SU2}

A multiple of the Pauli spin matrices  form one set of $J^{i}_{(2)}$s. We have 
\begin{equation} \label{Pauli1} J^{1} = \frac{1}{2}\pmatrix{0&&1 \cr 1 && 0}  \quad; \quad  J^{2} = \frac{1}{2}\pmatrix{0&&-i \cr i && 0}  \quad; \quad  J^{3} = \frac{1}{2}\pmatrix{1&&0 \cr 0 && -1}  \quad .
 \end{equation}
These $J^{i}_{(2)}$s have the following commutators 
\begin{equation} \label{Comm1} [J^{i}_{(2)},J^{j}_{(2)}]  = i \epsilon^{ijk} J^{k}_{(2)}  \quad,
 \end{equation}
where the commutator is defined by $[J^{i},J^{j}] \equiv$ $J^{i}J^{j}-J^{j}J^{i},$ matrix multiplication is understood, and $\epsilon^{ijk}$ is completely antisymmetric in $ijk$ with $\epsilon^{123}$ = 1. The commutation relations (\ref{Comm1}) form the Lie algebra of the group of rotations in 3-space.

The anticommutators of the above $J^{i}_{(2)}$s are
\begin{equation} \label{antiComm1} \{J^{i}_{(2)},J^{j}_{(2)}\}  = \frac{1}{2} \delta^{ij} {\mathbf{1}} \quad,
 \end{equation}
 where the anticommutator is defined by $\{J^{i},J^{j}\} \equiv$ $J^{i}J^{j}+J^{j}J^{i},$ $\delta^{ij}$ is the unit matrix, i.e. $\delta^{ij}$ = 0 for $i\neq j$ and $\delta^{ij}$ = 1 for $i = j,$ and ${\mathbf{1}}$ is the $2\times2$ unit matrix.  The summation convention is in force. It turns out that the anticommutator relations are important for the representation of boosts.

\section{Four Steps to Spacetime} \label{4steps}

\noindent Step 1. {\it{Two dimensional vector matrices and boost matrices.}}

Start by defining vector matrices ${V^{\mu}_{(2)}}_{ab}$, $\mu \in$ $\{1,2,3,4\},$  as follows,
\begin{equation} \label{Vs} V^{\mu}_{(2)} \equiv  \{c J^{1}_{(2)}, c J^{2}_{(2)}, c J^{3}_{(2)}, c^{4}{\mathbf{1}} \}  \quad,
 \end{equation}
where $c$ and $c^{4}$ are arbitrary scalars and ${\mathbf{1}}$ is the $2\times 2$ unit matrix. Also define three `boost matrices' ${K^{i}_{(2)}}_{ab},$ $i \in$ $\{1,2,3\},$ 
\begin{equation} \label{Ks} K^{i}_{(2)}  \equiv + i J^{i}_{(2)}   \quad.
 \end{equation}
There is a $\pm$ sign ambiguity associated with the definition of $i$ = $\sqrt{-1},$ so we could just as well have chosen $K^{i}_{(2)}$ to be $-iJ^{i}_{(2)}.$ This will be important in Step 2.

By the commutation relation (\ref{Comm1}), we see that the $V$s satisfy the following commutation relation, which is one of the Poincar\'{e} commutation relations,
\begin{equation} \label{VJ} [V^{\mu}_{(2)},J^{j}_{(2)}]  =  i \epsilon^{\mu jk} V^{k}_{(2)}  \quad ,
\end{equation}
where, for $\mu$ = 4, the antisymmetric symbol $\epsilon$ vanishes, $\epsilon^{4jk}$ = 0.

By multiplying (\ref{Comm1}) by 1, $i$ and $i^{2},$ one sees that the generators $J_{(2)}$ and $K_{(2)}$ obey the commutation relations
 \begin{equation} \label{Comm2} [J^{i},J^{j}]  = i \epsilon^{ijk} J^{k}  \quad ; \quad [J^{i},K^{j}]  = i \epsilon^{ijk} K^{k}  \quad ; \quad [K^{i},K^{j}]  =  -i \epsilon^{ijk} J^{k} \quad.
 \end{equation}
The 2-rep succeeds in replicating the Lorentz Lie algebra of angular momentum and boosts. 

The Poincar\'{e} algebra has commutators $[V^{i},K^{j}]$  that are symmetric in $i$ and $j.$ Both the  $V^{i}_{(2)}$s and the  $K^{j}_{(2)}$s are multiples of the  $J^{k}_{(2)}$s and we know that the commutators $[J^{i}_{(2)},J^{j}_{(2)}]$ of the $J$s are antisymmetric, not symmetric. Thus the 2-rep fails to give the Poincar\'{e} algebra of rotations, boosts, and translations.

\vspace{0.4cm}
\noindent Step 2. {\it{We move up to the $(2\oplus {2})$ rep of SU(2) to get symmetric $[V,K]$ commutation relations.}}
\vspace{0.01cm}

We get $ij$ symmetry in $[V^{i},K^{j}]$ by relating the commutators $[V,K]$ to the anticommutators $\{J,J\}$ in (\ref{antiComm1}). To implement the plan, introduce the $(2\oplus 2)$ SU(2) rep and define $4\times4$ matrices $J_{(2\oplus 2)}$s, $K_{(2\oplus 2)}$s, and $V_{(2\oplus 2)}$s as follows
 \begin{equation} \label{anti1} J^{i}_{(2\oplus 2)} = \pmatrix{J^{i}_{(2)} && 0 \cr 0 && J^{i}_{(2)}}  \, ; \, K^{i}_{(2\oplus 2)} = \pmatrix{+K^{i}_{(2)} && 0 \cr 0 && -K^{i}_{(2)}} \, ; \, V^{\mu}_{(2\oplus 2)} = \pmatrix{0 && {V^{\mu}_{+(2)}} \cr {V^{\mu}_{-(2)}} && 0} \, ,
 \end{equation}
where 
\begin{equation} \label{V+V-}  {V^{\mu}_{+(2)}} = \{c_{+} J^{1}_{(2)}, c_{+} J^{2}_{(2)}, c_{+} J^{3}_{(2)}, c^{4}_{+}{\mathbf{1}} \} \quad {\mathrm{and}}\quad {V^{\mu}_{-(2)}} = 
\{c_{-} J^{1}, c_{-} J^{2}_{(2)}, c_{-} J^{3}_{(2)}, c^{4}_{-}{\mathbf{1}} \} \quad.
\end{equation}
The constants $c_{+},$ $c^{4}_{+},$ $c_{-},$ $c^{4}_{-}$ are possibly different choices for the $c$s in (\ref{Vs}). 

The successful commutation relations from the 2-rep continue to work. One can verify the following 
 \begin{equation} \label{Comm3a} [V^{\mu}_{(2\oplus 2)},J^{j}_{(2\oplus 2)}]  =  i \epsilon^{\mu jk} V^{k}_{(2\oplus 2)} \quad
 \end{equation}
and 
 \begin{equation} \label{Comm2a} [J^{i}_{(2\oplus 2)},J^{j}_{(2\oplus 2)}]  = i \epsilon^{ijk} J^{k}_{(2\oplus 2)}  \quad ; \quad [J^{i}_{(2\oplus 2)},K^{j}_{(2\oplus 2)}]  = i \epsilon^{ijk} K^{k}_{(2\oplus 2)}  \quad ; \quad 
\end{equation}
$$[K^{i}_{(2\oplus 2)},K^{j}_{(2\oplus 2)}]  =  -i \epsilon^{ijk} J^{k}_{(2\oplus 2)} \quad. $$
 Thus the $J$s, $V$s and $K$s of the $(2\oplus {2})$ rep of SU(2) satisfy the $[J,J],$ $[J,K],$ $[K,K],$ $[V,J]$ commutation relations of the 2-rep in Step 1.
  
Now we find constants $c^{4}_{+}$ and $c^{4}_{-}$ in (\ref{V+V-}) such that the commutators $[V^{i}_{(2\oplus 2)},K^{j}_{(2\oplus 2)}]$ are sums of $V_{(2\oplus 2)}$s and are symmetric in $ij.$  The sign difference $K_{11}$ = $-K_{22}$ in (\ref{anti1}) turns blocks of the commutator $[V_{(2\oplus 2)},K_{(2\oplus 2)}]$ into anticommutators $\{V_{(2)},K_{(2)}\}.$ By (\ref{antiComm1}), (\ref{Ks}) and (\ref{anti1}), one finds that
\begin{equation} \label{Comm4}[V^{i}_{(2\oplus 2)},K^{j}_{(2\oplus 2)}] = \pmatrix{0&&-ic_{+}\{J^{i}_{(2)},J^{j}_{(2)}\} \cr +ic_{-}\{J^{i}_{(2)},J^{j}_{(2)}\}  &&0} = -i \delta_{ij}\pmatrix{0&& \frac{c_{+}}{2}{\mathbf{1}}\cr -\frac{c_{-}}{2}{\mathbf{1}}  &&0} \quad . 
\end{equation}
Comparing the result with the definition of $V^{4}_{(2\oplus2)}$ in (\ref{anti1}) and (\ref{V+V-}), the commutation relation, 
 \begin{equation} \label{Comm4a} [V^{i}_{(2\oplus 2)},K^{j}_{(2\oplus 2)}]  =  -i \delta_{ij} \alpha V^{4}_{(2\oplus 2)}  \quad,
 \end{equation}
is obeyed when $$2 \alpha c^{4}_{+} = c_{+} \quad {\mathrm{and}}\quad 2 \alpha c^{4}_{-} = -c_{-}  \quad,$$
and we have
\begin{equation} \label{anti3} V_{+}^{\mu} = c_{+}\{ J^{1}_{(2)},  J^{2}_{(2)},  J^{3}_{(2)}, {\mathbf{1}}/(2 \alpha) \} \quad {\mathrm{and}}\quad V_{-}^{\mu} = 
c_{-}\{ J^{1}_{(2)}, J^{2}_{(2)},  J^{3}_{(2)}, -{\mathbf{1}}/(2\alpha) \} \quad.
\end{equation}
In this way we obtain commutation relations $[V^{i},K^{j}]$ that are symmetric in $ij$ and are sums of $V_{(2\oplus 2)}$s. The $ij$ symmetry is characteristic of the Poincar\'{e} algebra.

 Also, by (\ref{Vs}), (\ref{anti1}) and (\ref{anti3}), one finds that 
\begin{equation} \label{V4Kj}	[V^{4}_{(2\oplus 2)},K^{j}_{(2\oplus 2)}]  = -\frac{i}{\alpha} V^{j}_{(2\oplus 2)} \quad .
  \end{equation}
By (\ref{Comm4a}) and (\ref{V4Kj}), the commutators $[V^{\mu},K^{j}]$ are sums of $V$s for any $\mu \in$ $\{1,2,3,4\}.$

Now we make momentum matrices. Momentum matrices are vector matrices $V^{\mu},$  $V^{\mu} \rightarrow$ $P^{\mu},$ that commute. Thus we have
 \begin{equation} \label{Comm5}  [P^{\mu}_{(2\oplus 2)},P^{\nu}_{(2\oplus 2)}]  = 0 \quad.
 \end{equation}
With $V \rightarrow P$ in  (\ref{anti1}), (\ref{anti3}) and (\ref{Comm5}), one sees that momentum matrices have $c_{+} c_{-}$ = 0 and one of the constants $c_{+}$ or $c_{-}$ must vanish.

Thus there are two versions of momentum matrices $P^{\mu}_{+(2\oplus 2)}$ and $P^{\mu}_{-(2\oplus 2)},$ each with just one nonzero off-diagonal block in this rep,
\begin{equation} \label{PP0} P^{\mu}_{+(2\oplus 2)} = \pmatrix{0 && {P^{\mu}_{+(2)}} \cr 0 && 0}    \quad {\mathrm{or}} \quad  P^{\mu}_{-(2\oplus 2)} = \pmatrix{0 && 0 \cr {P^{\mu}_{-(2)}} && 0} \quad,
\end{equation}
with $P^{\mu}_{+(2)}$ and $P^{\mu}_{-(2)}$ defined just like the $V_{+}^{\mu}$ and $V_{-}^{\mu}$ in (\ref{anti3}).

Combining the commutation relations (\ref{V4Kj}) and (\ref{Comm5}) with (\ref{Comm3a}) and (\ref{Comm2a}), one sees that the matrices $J^{j}_{(2\oplus 2)}$, $K^{j}_{(2\oplus 2)}$, $P^{\mu}_{\pm(2\oplus 2)}$  satisfy the Lie algebra of the Poincar\'{e} group:\cite{Poincare} 
\begin{equation} \label{PoincareJK} [J^{i},J^{j}]  = i\epsilon^{ijk} J^{k} \quad ; \quad [J^{i},K^{j}]  = i\epsilon^{ijk} K^{k} \quad ; \quad [K^{i},K^{j}]  = -i\epsilon^{ijk} J^{k} \quad
\end{equation}
\begin{equation} \label{PoincareVJK} [V^{\mu},J^{j}]  = i\epsilon^{\mu jk} V^{k} \quad ; \quad [V^{\mu},K^{j}]  = -i(\alpha \delta^{j \mu} \delta^{4\nu} + \frac{1}{\alpha}\delta^{j \nu} \delta^{4\mu}) V^{\nu} \quad 
\end{equation}
\begin{equation} \label{PoincarePP} [P^{\mu},P^{\nu}]  = 0 \quad ,
\end{equation}
where $\epsilon^{\mu jk}$ = 0 for $\mu$ = 4. It is understood that the momentum matrices $P^{\mu}$ satisfy the commutation relations (\ref{PoincareVJK}) for vector matrices $V^{\mu}.$

The Poincar\'{e} commutation relations (\ref{PoincareJK}), (\ref{PoincareVJK}), (\ref{PoincarePP}) can be rewritten so that the quantity $\alpha$ occurs only as a multiple of the time component $V^{4}_{(2\oplus 2)}.$ It expresses the ratio of space quantities, index 1, 2, or 3, to time quantities, index 4. One recognizes $\alpha$ as the speed of light, which we set equal to unity,
\begin{equation} \label{speedoflight} \alpha  = 1 \quad .
\end{equation}

{\it{Remark 2.1.}} Had we chosen to set $\alpha$ = $-1,$ the two spacetimes $\alpha$ = $\pm1$ would differ by a time inversion or a parity inversion. 

{\it{Remark 2.2.}} The $[V,K]$ commutation relation in (\ref{Comm4a}) and (\ref{PoincareVJK}) is a property of the boost matrices and vector matrices associated with the $(2\oplus {2})$ reducible rep of SU(2) that is not shared by the fundamental 2-dimensional irrep.

{\it{Remark 2.3.}} The fact that $[V,J]$ and $[V,K]$ are sums of $V$s does not make the set of $V$s an {\it{ideal}} because the commutator $[V,V]$ is not a sum of $V$s. One exception, of course, is when $[V,V]$ = 0. This specializes the $V$s as momentum matrices $P$ and, in that case, it is well known that the momentum matrices $P$ form an {\it{abelian ideal}} of the Poincar\'{e} algebra. 

\vspace{0.3cm}
\noindent Step 3. {\it{In this step the set of four vector matrices $V^{\mu}_{(2\oplus 2)}$ acts as a catalyst, carrying the Lorentz algebra of $J^{i}_{(2\oplus 2)}$s and $K^{i}_{(2\oplus 2)}$s from the $2\oplus2$-rep to the $3\oplus1$ = $4$-rep of spacetime. The vector matrices themselves are not carried to spacetime by the procedure.}}
\vspace{0.3cm}

Three facts underly the process in this step. First, the Lorentz Lie algebra of $J$s and $K$s is closed. Second, the commutators $[V,J]$ and $[V,K]$ are sums of $V$s. Third, the $V$s are off-diagonal.

Choose a $4\times4$ matrix $A^{i}$ and a matrix $B^{j}$ from the collection of $J^{i}_{(2\oplus 2)}$s and $K^{i}_{(2\oplus 2)}$s. One can write the commutation relation as follows,
\begin{equation} \label{ABC} [A^{i},B^{j}] = s^{ijk}_{(AB)} C^{k}   \quad,
\end{equation}
where the $C^{k}$s are  $J^{k}_{(2\oplus 2)}$s or $K^{k}_{(2\oplus 2)}$s and the coefficients $s^{ijk}_{(AB)}$ are found in (\ref{Comm2a}) or (\ref{PoincareJK}) for the particular choice of $A^{i}$ and $B^{j}$. 

By (\ref{Comm3a}),  (\ref{Comm4a}), (\ref{V4Kj}) and (\ref{PoincareVJK}), we have
\begin{equation} \label{VABC} [V^{\mu}_{(2\oplus 2)},A^{i}] = a^{\mu i \nu} V^{\nu}_{(2\oplus 2)} \quad ; \quad  [V^{\mu}_{(2\oplus 2)},B^{j}] = b^{\mu j \nu} V^{\nu}_{(2\oplus 2)} \quad ; \quad  [V^{\mu}_{(2\oplus 2)},C^{k}] = c^{\mu k \nu} V^{\nu}_{(2\oplus 2)}  \quad,
\end{equation}
where the coefficients $a$, $b$ and $c$ can be read from the equations for the particular choice of $A,B,C$ in (\ref{ABC}). From (\ref{VABC}), one can show that 
$$[V^{\rho}_{(2\oplus 2)},[A^{i},B^{k}]] =     \left( a^{\rho i \mu} b^{\mu k \nu} -  b^{\rho k \mu} a^{\mu i \nu} \right) V^{\nu}_{(2\oplus 2)}  \quad.
$$
Thus the vector matrices transfer the commutator of $A$ and $B$ to a commutator of $a$ and $b.$

Let $a^{i}$ be the $4\times4$ matrix with a $\rho \mu$th component equal to $a^{\rho i \mu}.$ Similarly define $b^{k}$ and $c^{n}.$ With (\ref{ABC}), the last equation implies that 
$$ \left[ a^{i},b^{k}\right]_{\rho \nu} V^{\nu}_{(2\oplus 2)} = s^{ikn}_{(AB)}c^{n}_{\rho \nu} V^{\nu}_{(2\oplus 2)} \quad.
$$
One cannot simply cancel the $V^{\nu}_{(2\oplus 2)}$s  because there is a sum over $\nu.$ And the four $V^{\nu}_{(2\oplus 2)}$s, $\nu \in$ $\{1,2,3,4\}$ do not form a linearly independent set of sixteen-component $4\times4$ matrices. 

But the $V^{\nu}_{(2\oplus 2)}$ are off-diagonal in this representation; see (\ref{anti1}). And the 2-dimensional blocks ${V^{\nu}_{+}}$ and ${V^{\nu}_{-}},$ each proportional to $\{ J^{1}_{(2)},  J^{2}_{(2)},  J^{3}_{(2)}, \pm{\mathbf{1}}/2 \},$ do form  linearly independent sets of $2\times2$ matrices. Then, by (\ref{anti1}), the preceding equation reduces to two equations involving $2\times2$ blocks and, by the linear independence of the four $2\times2$ matrices ${V^{\nu}_{+}}$ and ${V^{\nu}_{-}},$ we have shown that 
\begin{equation} \label{theorem3} \left[ a^{i},b^{k}\right] = s^{ikn}_{(AB)}c^{n} \quad .
\end{equation}
The coefficients $a^{i},$ $b^{k}$ and $c^{n}$ obey the same commutation relation (\ref{ABC}) as $A^{i},$ $B^{k}$ and $C^{n}.$ 

By (\ref{theorem3}), the coefficients in (\ref{Comm3a}),  (\ref{Comm4a}), and (\ref{V4Kj}) form angular momentum matrices ${J^{i}_{(4)}}_{\rho \mu}$ and boost matrices ${K^{i}_{(4)}}_{\rho \mu},$
 \begin{equation} \label{4JK} {J^{i}_{(4)}}_{\rho \mu}  =  i\epsilon^{\rho i \mu} \quad ; \quad {K^{i}_{(4)}}_{\alpha \beta}  =  -i \left(\delta^{i}_{\alpha} \delta^{4}_{\beta} +\delta^{i}_{\beta} \delta^{4}_{\alpha}\right)  \quad.
 \end{equation}
By  (\ref{theorem3}) or directly from (\ref{4JK}), these ${J^{i}_{(4)}}$s and ${K^{i}_{(4)}}$s satisfy the Lorentz Lie algebra,
 \begin{equation} \label{Comm2b} [J^{i}_{(4)},J^{j}_{(4)}]  = i\epsilon^{ijk} J^{k}_{(4)}  \quad ; \quad [J^{i}_{(4)},K^{j}_{(4)}]  = i\epsilon^{ijk} K^{k}_{(4)}  \quad ; \quad 
[K^{i}_{(4)},K^{j}_{(4)}]  =  -i\epsilon^{ijk} K^{k}_{(4)} \quad, 
\end{equation}
where `$(4)$' is short for $(3\oplus 1)$ and indicates the `4-vector' rep. The ${J^{i}_{(4)}}$s generate rotations in the three-space of spacetime. Note that the ${J^{i}_{(4)}}$s make up the adjoint rep of SU(2). The ${K^{i}_{(4)}}$s generate boosts in spacetime. 

{\it{Remark 3.1.}} By tracing the origins of the angular momentum matrices ${J^{i}_{(4)}}$ in (\ref{4JK}) through (\ref{VABC}) to (\ref{anti3}) to (\ref{Comm3a}) to (\ref{Comm1}), one sees that the ${J^{i}_{(4)}}$s are based on the antisymmetric structure constants of the commutators $[J_{(2)},J_{(2)}]$ of the angular momentum matrices $J^{i}_{(2)}$ in the fundamental 2-rep of SU(2) in (\ref{Comm1}). 

{\it{Remark 3.2.}} Tracing the origins of the boost matrices ${K^{i}_{(4)}}$ (\ref{4JK}), we follow the ${K^{i}_{(4)}}$s from (\ref{VABC}) to  (\ref{Comm4a}) and (\ref{V4Kj}) with (\ref{Comm3a}) back to (\ref{antiComm1}). One sees that the boost matrices ${K^{i}_{(4)}}$ evolve from the anticommutators $\{J_{(2)},J_{(2)}\}$ of the  angular momentum matrices in the fundamental 2-rep of SU(2) in (\ref{antiComm1}).

\vspace{0.3cm}
\noindent Step 4. {\it{Invariants. The rotation invariance of the square of the distance ${x^{i}}^{2}$ and time $x^{4}$ depend on fundamental properties. The invariance of ${x^{i}}^{2}-{x^{4}}^{2}$ under boosts has weaker support. }}
\vspace{0.3cm}

In this step we use the Lorentz transformation $D_{(4)}(\theta,\phi)$ = $\exp{(i\phi_{i}K^{i}_{(4)})}\exp{(i\theta_{i}J^{i}_{(4)})},$ the 4-vector transformation matrix for a rotation through angle $\theta$ = $\{\theta_{1},\theta_{2},\theta_{3}\}$ followed by a boost through $\phi$ = $\{\phi_{1},\phi_{2},\phi_{3}\}.$

The invariance of distance under rotations can be shown by considering an infinitesimal rotation. For small $\theta,$ $\mid \theta_{i} \mid \ll$ 1, and with a scalar $x^{\mu}$ so that components commute, $x^{i}x^{j}$ = $x^{j}x^{i},$ one finds that
$${{x^{ \prime}}^{\, i}}^{2} = \left(\delta_{ij} + i  \theta_{k} {J^{k}_{(4)}}_{ij}\right)x^{j} \left(\delta_{il} + i  \theta_{m} {J^{m}_{(4)}}_{il}\right)x^{l}  = {x^{i}}^{2} + 2i \theta_{k}{J^{k}_{(4)}}_{ij} x^{i}x^{j} =  {x^{i}}^{2}\quad,
$$
where terms of order $\theta^{2}$ are dropped, ${x^{i}}^{2}$ = ${x^{1}}^{2}+{x^{2}}^{2}+{x^{3}}^{2}$ and the prime indicates the rotated 4-vector ${x^{ \prime}}$ = $D_{(4)}( \theta,0)x.$  Invariance follows for finite angles also. 

The invariance follows because ${J^{i}_{(4)}}_{ij}$ is antisymmetric in $ij$ making ${J^{k}_{(4)}}_{ij} x^{i}x^{j}$ vanish. By (\ref{4JK}) ${J^{i}_{(4)}}_{ij}$ is antisymmetric because the structure constants of SU(2) in (\ref{Comm1}) are antisymmetric. This is a fundamental property.

Since the time components, index = 4, of the ${J^{i}_{(4)}}$s vanish, ${J^{i}_{(4)}}_{\alpha4}$ = ${J^{i}_{(4)}}_{4\beta}$ = $i\epsilon^{i \alpha 4}$ = $i\epsilon^{i 4 \beta}$ = 0, one finds that the time component $x^{4}$ is invariant. Collecting rotation invariants, we have
  \begin{equation} \label{x2ROTa}  {{x^{ \prime}}^{\, i}}^{2} = {x^{i}}^{2} \quad ; \quad {{x^{ \prime}}^{\, 4}} = {x^{4}} \quad,
 \end{equation}
where the prime indicates the rotated 4-vector ${x^{ \prime}}$ = $D_{(4)}(\theta,0)x.$

In contrast to the ${J^{i}_{(4)}}$s, the spacetime boost generators are symmetric, ${K^{i}_{(4)}}_{\alpha \beta}$ = $+{K^{i}_{(4)}}_{\beta \alpha}$ and the time components of ${K^{i}_{(4)}}$ do not vanish. Thus neither distance nor time is preserved under boosts. 

With boosts, the invariant is the difference of the square of the distance and the square of the time, i.e. the square of the spacetime `interval'. For small $\phi,$ $\mid \phi_{i} \mid \ll$ 1, one finds that 
$$ {{x^{\prime \prime}}^{\, i}}^{2}-{{x^{\prime \prime}}^{\, 4}}^{2} = \hspace{5cm} \quad $$
 \begin{equation} \label{x2BOOST} = \left(\delta_{j\sigma} + i \phi_{n} {K^{n}_{(4)}}_{j\sigma}\right)x^{\sigma}\left(\delta_{j\rho} + i \phi_{m} {K^{m}_{(4)}}_{j\rho}\right)x^{\rho} - \left(\delta_{4\sigma} + i \phi_{n} {K^{n}_{(4)}}_{4\sigma}\right)x^{\sigma}\left(\delta_{4\rho} + i \phi_{m} {K^{m}_{(4)}}_{4\rho}\right)x^{\rho} = 
 \end{equation}
$$=  {x^{i}}^{2} -{x^{4}}^{2}+ 2i\phi_{n}\left[{K^{n}_{(4)}}_{j\sigma} \, x^{j}x^{\sigma} - {K^{n}_{(4)}}_{4\sigma} \, x^{4}x^{\sigma}\right]=  {x^{i}}^{2}-{x^{4}}^{2} \quad,$$
where terms of order $\phi^{2}$ are dropped. The last equality follows from the particular values (\ref{4JK}) of the boost matrices ${K^{i}_{(4)}}.$ Thus the square of the spacetime interval, ${x^{i}}^{2}-{x^{4}}^{2},$ is invariant under boosts,
  \begin{equation} \label{x2BOOSTa} {{x^{\prime \prime}}^{\, i}}^{2}-{{x^{\prime \prime}}^{\, 4}}^{2}  = {x^{i}}^{2}-{x^{4}}^{2} \quad,
 \end{equation}
where the double prime indicates the boosted rotated 4-vector ${x^{\prime \prime}}$ = $D_{(4)}(\theta,\phi)x.$

Making $\left[{K^{n}_{(4)}}_{j\sigma} \, x^{j}x^{\sigma} - {K^{n}_{(4)}}_{4\sigma} \, x^{4}x^{\sigma}\right]$  vanish in (\ref{x2BOOST}) depends on the particular values of the components of the boost matrices ${K^{i}_{(4)}}.$ The ${K^{i}_{(4)}}$s in (\ref{4JK}) are combinations of delta functions. The delta functions derive from the delta function in (\ref{antiComm1}), the anticommutator of the $J_{(2)}$s in the fundamental 2-rep of SU(2). For general SU($N$), $N \geq 3,$ the anticommutators of the $J_{(N)}$s contain additional terms. Thus SU(2) has special, simple anticommutators and the invariance of the spacetime interval under boosts is a direct consequence of the simplicity of SU(2).

Finally, the translation invariance of coordinate differences occurs because the coordinate differences themselves are invariant under translations,
  \begin{equation} \label{trans}  {x^{\prime \prime \prime}}^{\alpha}  =  ({x_{1}}^{\alpha}+a^{\alpha}) - ({x_{0}}^{\alpha}+a^{\alpha}) = {x_{1}}^{\alpha} - {x_{0}}^{\alpha} =  x^{\alpha} \quad,
 \end{equation}
where the triple prime indicates a translation through the displacement $a^{\alpha}.$ 

The Poincar\'{e} transformation of the coordinates $x^{\mu}$ through a rotation followed by a boost followed by a translation can be represented by first appending a one to the coordinates. One has
\begin{equation} \label{RBT1} 
  \pmatrix{[D_{(4)}(\theta,\phi)]^{\mu}_{\nu}&&a^{\mu} \cr 0 && 1} \pmatrix{x^{\nu} \cr 1} = \pmatrix{x^{\prime \, \mu} +a^{\mu} \cr 1}\quad .
 \end{equation}
 The prime indicates $x$ has been rotated then boosted, $x^{\prime}$ = $D_{(4)}(\theta,\phi)\,x. $
 
 For rotations and boosts generated with the ${J^{i}_{(4)}}$s and ${K^{i}_{(4)}}$s and with $x$  a coordinate difference, we have shown that the square of the spacetime interval ${x^{i}}^{2}-{x^{4}}^{2}$ is invariant under rotations, boosts, and translations.

 {\it{Remark 4.1.}} The device used in representing translations, i.e. appending the one to $x^{\mu},$ is a special case. In general, one can append a second order tensor $T^{\mu \nu}$ to the vector $x^{\mu}$ and define a Poincar\'{e} transformation to the combination.~\cite{S2} In (\ref{RBT1}), the tensor is a scalar. 

The device is related to generalized Dirac matrices. Such matrices are needed to be combined with the gradient in linear relativistic wave equations for wave functions with general spins.\cite{Lyubarskii} The spin in the case of (\ref{RBT1}) is $(1/2,1/2)\oplus(0,0).$ The discussion lies beyond the scope of the present article and may be treated elsewhere.

{\it{Remark 4.2.}} Notice that the invariance of distance in 3-space under rotations is {\it{robust}} because it depends mainly on the antisymmetry of the ${J^{i}_{(4)}}$s which ultimately depends on the antisymmetry of structure constants. The invariance of the square of the spacetime interval, ${x^{i}}^{2}-{x^{4}}^{2},$ under boosts is {\it{not robust}} since the invariance depends on the simplicity of the anticommutators of the generators $J_{(2)}$ in the fundamental 2-rep of SU(2).

\pagebreak


\section{Exercises and Problems} \label{Pb}

\noindent Exercises and problems are provided online. Answers and hints are hidden as remarks (\%) in the LaTeX source document available from the physics arXiv. Download the source from the physics arXiv, remove the {\%}s from this section, and compile the LaTeX. 

\vspace{0.3cm}
\noindent The Pauli matrices are
$$ \sigma^{1} = \pmatrix{0&&1 \cr 1 && 0} \quad  \sigma^{2} = \pmatrix{0&&-i \cr i && 0} \quad \sigma^{3} = \pmatrix{1&&0 \cr 0 && -1} \quad \sigma^{4} = \pmatrix{1&&0 \cr 0 && 1} \quad .$$
Let $\gamma^{\mu}$ be the following set of Dirac gamma matrices,
$$  \gamma^{i} =  -i\pmatrix{0&& \sigma^{i} \cr -\sigma^{i} && 0} \quad \gamma^{4} =  -i\pmatrix{0&& \sigma^{4} \cr \sigma^{4} && 0} \quad. $$

\vspace{0.3cm}
\noindent 1. Show that $$ -\det{(x^{\mu} \sigma^{\mu})} = {x^{i}}^{2} - {x^{4}}^{2} \quad,$$ where the summation convention applies and $\mu \in$ $\{1,2,3,4\}$ and $i \in$ $\{1,2,3\}.$


\vspace{0.3cm}
\noindent 2. Show that $J^{i}$ = $\sigma^{i}/2$ satisfy the commutation and anticommutation relations (\ref{Comm1}) and (\ref{antiComm1}). 


\vspace{0.3cm}
\noindent 3. The Dirac gamma matrices are vector matrices.  Find values of $c_{+}$ and $c_{-}$ in (\ref{anti1}) and (\ref{anti3}) such that $V^{\mu}_{(2\oplus 2)}$ = $\gamma^{\mu}.$


\vspace{0.3cm}
\noindent 4. {\it{Parity - momentum matrix relationship.}} Define $\gamma^{5}$ by $\gamma^{5} \equiv$ $-i \gamma^{4}\gamma^{1}\gamma^{2}\gamma^{3}.$ Given a vector matrix $V^{\mu}_{(2 \oplus2)},$ show that $$P^{\mu}_{+} =  \frac{1}{2}({\mathbf{1}} + \gamma^{5}) V^{\mu}_{(2 \oplus2)} \quad {\mathrm{and}} \quad P^{\mu}_{-} =  \frac{1}{2}({\mathbf{1}} - \gamma^{5}) V^{\mu}_{(2 \oplus2)} \quad$$ are momentum matrices, i.e. they obey the $[V,J],$ $[V,K],$ and $[P,P]$ commutation relations (\ref{PoincareVJK}) and (\ref{PoincarePP}) with the $J^{i}_{(2\oplus2)}$s and the $K^{i}_{(2\oplus2)}$s. [Note: In quantum mechanics, the matrices $({\mathbf{1}} \pm \gamma^{5})/2$ are used to project out Left(+) and Right($-$) handed parts of spin 1/2 fields.]

\vspace{0.3cm}
\noindent 5. Find the generators $J^{i},$ $K^{i},$ $P^{\mu}$ of the Poincar\'{e} transformation in (\ref{RBT1}). Show the matrices obey the Poincar\'{e} algebra (\ref{PoincareJK}), (\ref{PoincareVJK}), (\ref{PoincarePP}).


\vspace{0.3cm}
\noindent 6. Given the $[V,J]$ and $[V,K]$ Poincar\'{e} commutation relations (\ref{PoincareVJK}) for general rep $J$s, $K$s, $V$s and the spacetime 4-vector rep generators ${J^{i}_{(4)}}_{\rho \mu}$  =  $i\epsilon^{\rho i \mu}$ and ${K^{i}_{(4)}}_{\alpha \beta}$  =  $-i (\delta^{i}_{\alpha} \delta^{4}_{\beta}$ + $\delta^{i}_{\beta} \delta^{4}_{\alpha})$ in (\ref{4JK}), show that $$DV^{\mu}D^{-1} = \Lambda^{\mu}_{\nu}V^{\nu} \quad.$$ In this expression, $\Lambda$ = $\Lambda(\theta,\phi)$ = $\exp{(i\phi_{i}K^{i}_{(4)})}\exp{(i\theta_{i}J^{i}_{(4)})}$ is the 4-vector transformation matrix generated by $J^{i}_{(4)}$ and $K^{i}_{(4)}$ for a rotation through the angle $\theta$ = $\{\theta^{1},\theta^{2},\theta^{3}\}$ followed by a boost for $\phi$ = $\{\phi^{1},\phi^{2},\phi^{3}\}.$ The matrix $D$ = $D(\theta,\phi)$ = $\exp{(i\phi_{i}K^{i})}\exp{(i\theta_{i}J^{i})}$ represents the same transformation generated by the $J$s and $K$s.


\vspace{0.3cm}
\noindent 7. Compare the $[J,J]$  commutator and the $\{J,J\}$ anticommutator relations for SU(2) with those for SU(3).  Can you determine the eight `spacetime' `angular momentum' matrices $J^{i}_{(9)}$ from the $[J,J]$ SU(3) commutator relations? What about getting the the eight $K^{i}_{(9)}$s from the anticommutator relations?



\begin{thebibliography}{9}


\bibitem{Tung} See, for example, Wu-Ki Tung, {\it{Group Theory in Physics}} (World Scientific, Singapore, 1985), Chapter 8.


\bibitem{SU2} See, e.g., J. Matthews and R.L. Walker, {\it{Mathematical Methods of Physics}} (W.A. Benjamin, New York, 1965): Chapter 16, Sections 6 and 7. 


\bibitem{Poincare}  See, e.g., S. Weinberg, {\it{The Quantum Theory of Fields}} (Cambridge Univeristy Press, Cambridge, 1995): Chapter 2, Section 4.



\bibitem{S2} R. Shurtleff, article in the on-line physics arxiv: {\it{Poincare Connections in Flat Spacetime}}, 	http://arXiv:gr-qc/0502021v2 .


\bibitem{Lyubarskii} G.Ya. Lyubarskii, {\it{The Application of Group Theory in Physics}}, translated from the Russian by Steven Dedijer (Pergamon Press, Oxford, 1960): Section 74.
 
\end{thebibliography}
\end{document}